\documentclass[12pt]{article}
\usepackage{graphics}
\begin{document}

\large
\begin{enumerate}
\item 24 March 1997 (final accepted version)
\item Effect of the orthorhombic distortion
on the magneto-optical properties of SrRuO$_3$
\item I. V. Solovyev
\item
Joint Research Center for Atom Technology, Angstrom Technology Partnership,
1-1-4 Higashi, Tsukuba, Ibaraki 305, Japan\\
\item Abstract \par \hspace{1cm}
It is argued that the non-collinear orbital magnetism
accompanying the orthorhombic distortion is an important
ingredient which should be taken into consideration
for making promising materials with the best
magneto-optical characteristics on the basis of SrRuO$_3$.
\item {\it keywords:}
Magneto-optical properties,
Canted ferromagnetism,\\
Band calculations,
Perovskite structure
\item Corresponding author.\\
I. V. Solovyev\\
JRCAT-ATP, c/o NAIR\\
1-1-4 Higashi \\
Tsukuba, Ibaraki 305\\
JAPAN\\
Fax:+81-298-54-2788\\ E-mail: igor@jrcat.or.jp
\end{enumerate}

\newpage

  Besides other interesting properties,
SrRuO$_3$ is considered as one of rare examples of 4d itinerant
magnets and a good candidate for applications in magneto-optical devices
because of the large spin-orbit interaction (SOI) at the
Ru sites \cite{Klein}.
The perspectives seem to be even more favorable after recent
band structure calculations predicting nearly half-metallic behavior
for SrRuO$_3$ in the actual orthorhombic $D^{16}_{2h}$ structure
\cite{Singh}. Unfortunately, the magnetic ordering compatible with
the space group $D^{16}_{2h}$ implies a non-collinearity
\cite{Treves}, which can be especially large for the
orbital counterpart \cite{Igor_MO}. It means, although the spin
magnetic moments ($M_S$) are ordered ferromagnetically, the
orbital magnetic moments ($M_L$) can be canted off the
spin directions and show an appreciable antiferromagnetic
component. The latter is definitely a destructive factor for the
magneto-optical characteristics of SrRuO$_3$.
The situation is just opposite to that considered in Ref.\cite{Igor_MO},
where essentially non-collinear orbital magnetic ordering was the
source of a pronounced optical nonreciprocity in the series of
La$M$O$_3$ perovskites displaying only weak spin ferromagnetism.

  In the present work we investigate the orbital
non-collinearity and its consequences
on the magneto-optical properties of SrRuO$_3$
using the band structure calculations in the local-spin-density
approximation.
Structural parameters were taken from
Ref. \cite{Shikano}.
In our numerical procedure
we align the
spin magnetic moments
sequentially along the principal
${\bf a}$, ${\bf b}$ and ${\bf c}$ directions of the orthorhombic cell
and evaluate the responding band structure
after including SOI as a pseudoperturbation in
the ASA-LMTO method \cite{LMTO}.
Then, the interband optical conductivity can be calculated using
Eq.(5.1) of the work \cite{Wang}.
Details can be found elsewhere
\cite{Igor_MO}.

  In the cubic perovskite structure, $M_S$ at the Ru site is
$1.220 \mu_B$, whereas $M_L$ is almost negligible
($-0.005 \mu_B$) due to the large cancelation of contributions
with different spin quantum numbers.
The effect of the orthorhombic distortion, which is
accompanied by substantial rotations of RuO$_6$ octahedra
relative to each other, is twofold.
As the Ru$-$O$-$Ru bond angles are reduced \cite{Shikano},
the 4d-band is narrowed \cite{Singh} and the spin polarization
is increased ($M_S=1.469\mu_B$). These factors enhance
the local orbital moments (Table \ref{tab.orb}).
On the other hand, the orientational modulation of the
RuO$_6$ octahedra causes analogous changes in the directions of the
orbital moments \cite{Igor_MO}. Indeed,
while initial arrangement
of the spin magnetic moments was ferromagnetic and totally collinear,
the obtained distribution of the orbital magnetic moments
largely deviates from the ferromagnetic one
(Table \ref{tab.orb})
and, depending on the direction of the spin magnetization,
belongs to one of the four
types compatible with the space group $D^{16}_{2h}$ \cite{Treves}.

\begin{figure}[h]
\centering \noindent
\resizebox{10cm}{!}{\includegraphics{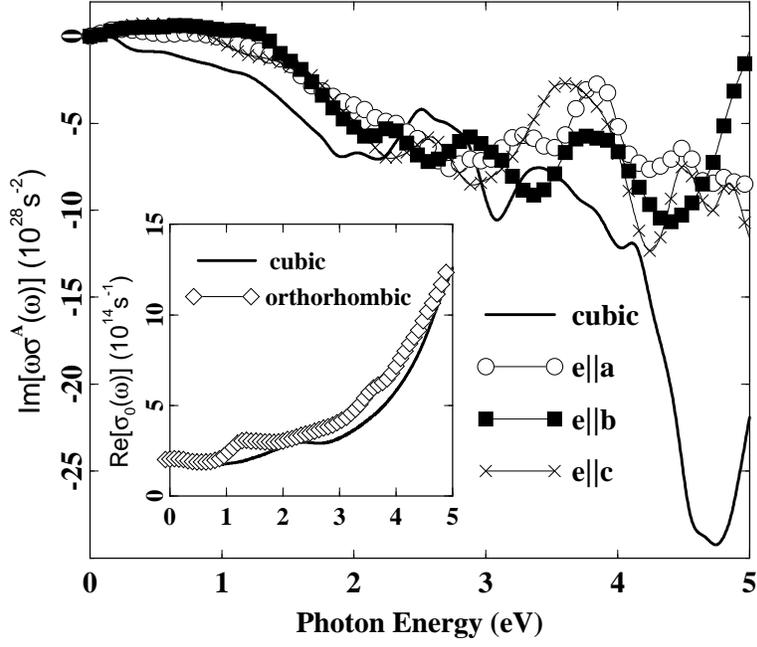}}
\caption{Antisymmetric ($\sigma^A$) and diagonal ($\sigma_0$)
         parts of the interband optical conductivity
         calculated in the
         cubic perovskite and the actual orthorhombic structure
         of SrRuO$_3$ for three possible directions of the
         ferromagnetic spin moments in the orthorhombic cell.
         The phenomenological inverse relaxation time of 10mRy
         was used for the broadening.}
\label{fig.OC}
\end{figure}

\begin{figure}[h]
\centering \noindent
\resizebox{10cm}{!}{\includegraphics{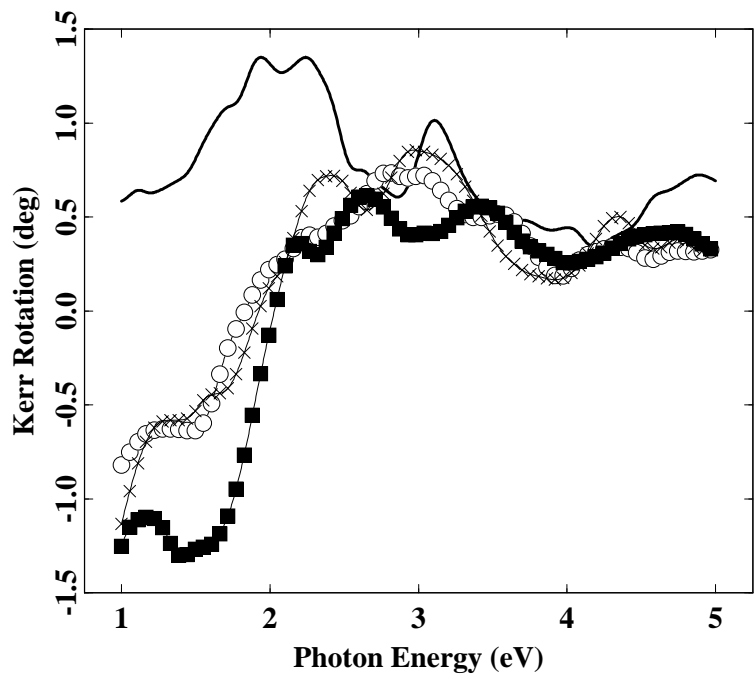}} \caption{Polar
Kerr rotation corresponding to the
         optical conductivities shown in Fig.1.
         No intraband Drude contributions have been added.
         Therefore, the infrared region of the spectra is not
         shown.}
\label{fig.Kerr}
\end{figure}

  Similar behavior is expected for the magneto-optical properties.
For any direction of the ferromagnetic moment $\alpha$,
the antisymmetric part of the conductivity tensor
$\sigma^A(\omega) = (1/2)
\sum_{\beta \gamma} \varepsilon_{\alpha \beta \gamma}
\sigma_{\beta \gamma}(\omega)$
is strongly suppressed in the region
corresponding to the charge transfer excitations
(around 4.5 eV in Fig.\ref{fig.OC}) when the orthorhombic
distortion is taken into account, whereas the diagonal part
$\sigma_0(\omega)=(1/3) \sum_\beta \sigma_{\beta \beta}
(\omega)$ undergoes only small changes
Thus, the nonreciprocal optical rotation
being
proportional to $\sigma^A(\omega)$ can be
renormalized dramatically in the orthorhombic
SrRuO$_3$. An example of the polar Kerr
rotation, estimated as
\begin{displaymath}
\theta_K(\omega) \approx {\rm Re} [ \frac{- \sigma^A(\omega)}
{\sigma_0(\omega) \sqrt{1+\sigma_0(\omega) 4 \pi {\it i}/ \omega}}],
\end{displaymath}
is shown in Fig.\ref{fig.Kerr}
(we found that even in the orthorhombically distorted SrRuO$_3$,
the anisotropy of the diagonal conductivity is small and
in fact $\sigma_{\beta \beta} (\omega) \approx \sigma_0(\omega)$).
The renormalization will presumably be even more pronounced due to
essential
non-collinearity of the spin arrangement in this compound expected
from the analysis of interatomic magnetic interactions
\cite{Igor_NC}, which implies a self-consistent treatment
for the directions of the spin magnetic moments
beyond the scopes of the
present calculations. \\

  The author is grateful to K. Terakura, Y. Suzuki
and Y. Yokoyama for valuable discussions.
The work
is partly supported by New Energy and Industrial
Technology Development Organization (NEDO).

\newpage

\begin{table}[h]
\caption{Three possible directions of the ferromagnetic spin axis in the
         orthorhombic cell ($S$),
         orbital magnetic ordering induced by SOI ($L$),
         local
         orbital moment at Ru site ($M_L$) and
         deviation of the orbital moment from the
         ferromagnetic axis ($\Psi$).
         The non-collinear orbital ordering is classified by the type
         (antiferro-A, C, G and ferro-F) for each ${\bf a}$, ${\bf b}$ and
         ${\bf c}$ projection of the magnetic moment.}
\label{tab.orb}
\begin{center}
\begin{tabular}{cccc}\hline
$S$                & $L$                                 & $M_L$ ($\mu_B$) & $\Psi$ ($^\circ$)   \\\hline
${\bf e}\|{\bf a}$ & F$_{\bf a}-$C$_{\bf b}-$G$_{\bf c}$ & 0.061  & 34.5     \\
${\bf e}\|{\bf b}$ & C$_{\bf a}-$F$_{\bf b}-$A$_{\bf c}$ & 0.054  & 37.9     \\
${\bf e}\|{\bf c}$ & G$_{\bf a}-$A$_{\bf b}-$C$_{\bf c}$ & 0.069  & 40.9     \\\hline
\end{tabular}
\end{center}
\end{table}

\newpage

\end{document}